\documentclass[aps,pra,twocolumn,floatfix]{revtex4}
\usepackage{graphicx}
\usepackage{amsmath}

\begin{document}

\title{Magic ratio of window width to grating period for Van der Waals potential measurements using material gratings}
\author{Vincent P. A. Lonij}
\author{William F. Holmgren}
\author{Alexander D. Cronin}
\affiliation{University of Arizona, Department of Physics, Tucson, AZ 85721}
\begin{abstract}

We report improved precision measurements of the Van der Waals potential strength ($C_3$) for Na atoms and a silicon-nitride (SiN$_x$) surface. We studied diffraction from nano-fabricated gratings with a particular ``magic'' open-fraction that allows us to determine $C_3$ without the need for separate measurements of the width of the grating openings. Therefore, finding the magic open-fraction improves the precision of $C_3$ measurements. The same effect is demonstrated for a grating with an arbitrary  open-fraction by rotating it to a particular ``magic'' angle, yielding $C_3=3.42\pm 0.19 \textrm{ eV} \mathring{\textrm{A}}^3$ for Na and a SiN$_x$ surface. This precision is sufficient to detect a change in $C_3$ due to a thin metal coating on the grating surface. We discuss the contribution to $C_3$ of core electrons and edge effects.

\end{abstract}
\maketitle

\section{Introduction}

Van der Waals and Casimir-Polder potentials are the dominant interactions between charge-neutral objects at nano- to micrometer length scales. As such they have attracted considerable interest in the field of microtechnology, as well as the field of quantum-gravity, where some theories predict deviations from Newtonian potentials at very short length scales \cite{Dimopoulos:2003ty}. The dependence of these potentials on the geometry and dielectric function of surfaces is only partially understood. In the present paper we report measurements of sufficient precision to study these effects.

Over the past decade, Van der Waals (VdW) potentials between atoms and surfaces have been measured using diffraction 
from nano-fabricated gratings \cite{Grisenti:1999mz,Bruhl:2002ty,Perreault:2005kx}, quantum reflection from surfaces \cite{Druzhinina:2003rw,Mohapatra:2006eu}, and 
spectroscopy of Cs atoms in nano-cells \cite{Fichet:2007yg}. The most accurate of these reported an uncertainty of 15\% \cite{Mohapatra:2006eu}. 

In this paper we report the VdW potential strength ($C_3$) for Na atoms and a silicon-nitride (SiN$_x$) surface with a precision of $5\%$ by studying diffraction of an atom beam from a nano-grating. We applied this improved precision to show that we can detect a change in $C_3$ due to a thin layer of metal deposited on the grating. In addition we will discuss the effect of edges on the atom-surface potential and we will discuss some of the difficulties in calculating the VdW potential for real systems.

The dominant source of uncertainty in $C_3$ in previous diffraction experiments was imprecise knowledge of the grating's geometric parameters.  Perreault \textit{et al.} \cite{Perreault:2005kx} reported an uncertainty in $C_3$ of 25\%  due to an uncertainty in the window width ($w$) of only 1 nm (when $w$= 50 nm). This correlation between $C_3$ and $w$ is the main problem that we have overcome by finding the ``magic'' open-fraction. We can now measure $C_3$ independently of $w$.

Perreault \textit{et al.} determined $w$ by SEM imaging; conventional imaging techniques however cannot easily improve these measurements. SEM/ TEM imaging is hindered by charging and image charge effects, while STM/AFM images show a convolution of the sample and the unknown tip shape.
The experiment described in this paper allows us to determine the geometric parameters in addition to $C_3$.

\begin{figure}
\begin{center}
\includegraphics[width=8.5cm]{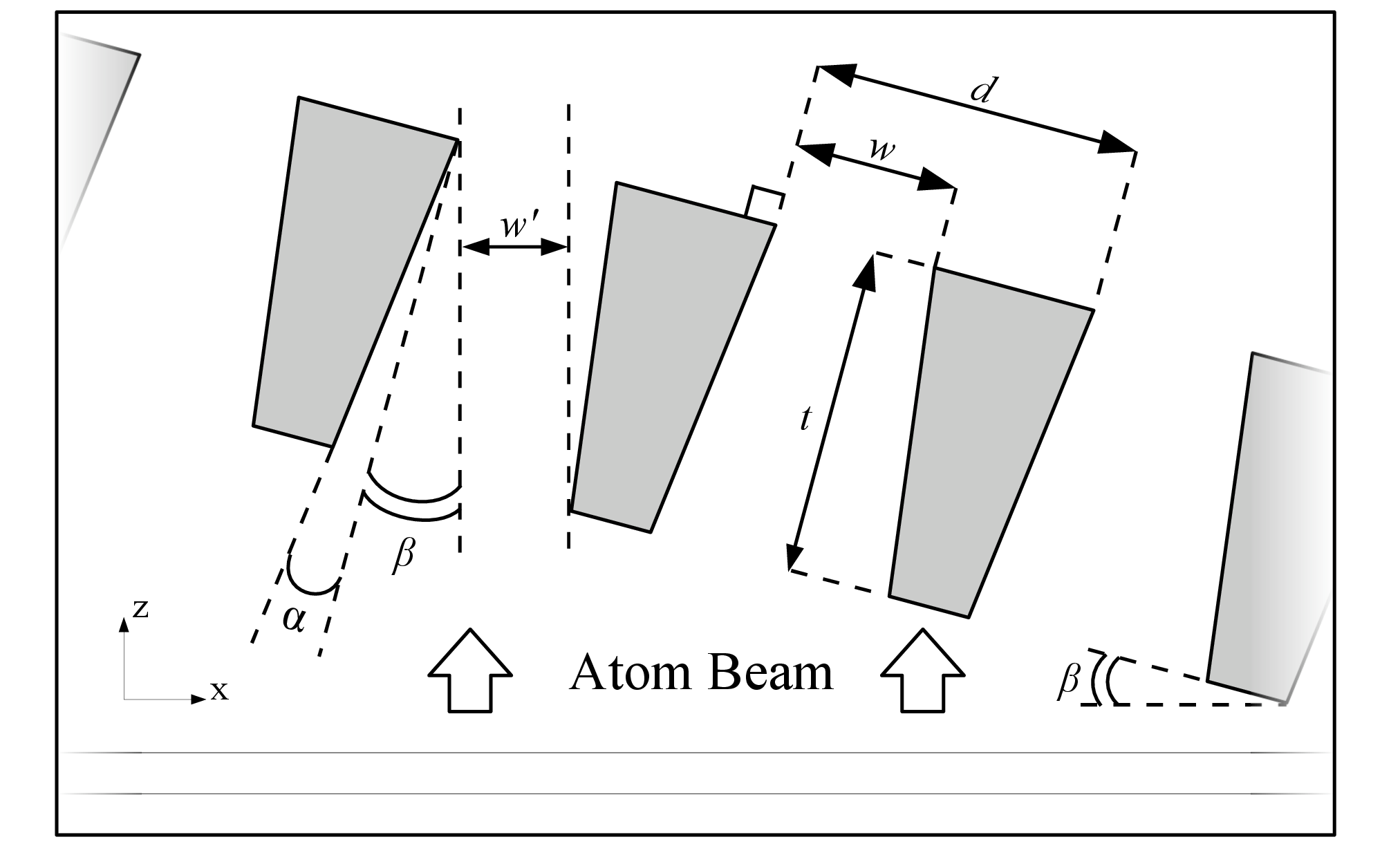}
\caption{\label{fig:gr_dr} A schematic representation of the grating geometry. The grating bars have a trapezoidal cross-section, 
with a wedge angle $\alpha$. The grating can be rotated around an axis parallel to the grating bars ($y$) by an 
angle $\beta$. The period of the grating ($d$) is 100 nm. The width of the grating windows ($w$) varies for different 
gratings between 40 and 70 nm. The thickness of the bars ($t$) is about 120 nm. $w'$ is the projected widow width.}
\end{center}
\end{figure}

The reason it was difficult to determine either $C_3$ or $w$ from a study of diffraction alone is that the effect of a small increase in $C_3$ is usually very similar to the effect of a small decrease in the window width. In this paper we explain the origin of this correlation and we will show how the correlation disappears for certain special values of the open-fraction $w/d$. Gratings with this magic open-fraction enable us to determine $C_3$ and $w$ independently from each other, thus eliminating the effect that limited the precision of previous experiments. 

We have experimentally demonstrated this by repeatedly coating a grating with metal until the desired $w_\textrm{magic}$ was reached (Section II). In the study of atom-surface interactions however, it is undesirable to contaminate the surface. We therefore developed a similar method wherein the grating is rotated by an angle, $\beta_m$, to change the projected open-fraction such that the $C_3$ - $w$ correlation vanishes (Section III).

\begin{figure}
\begin{center}
\includegraphics[width=8.5cm]{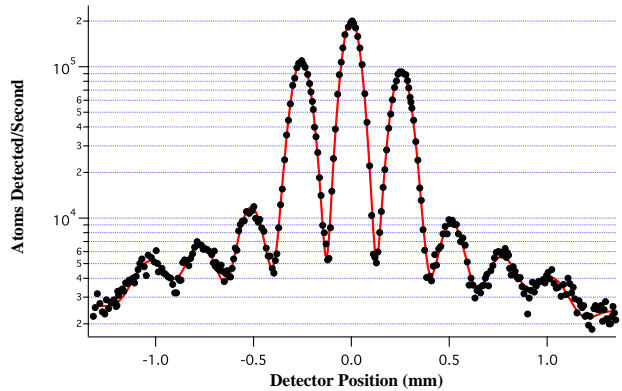}
\caption{(Color online) Atom flux as a function of detector position in the far field, 2.43 m from a rotated diffraction grating. Note that at non-normal 
incidence, the positive orders differ in intensity from the negative orders. This asymmetry is well described by theory \cite{Cronin:2004yq}.}
\label{fig:raw_data}
\end{center}
\end{figure}

\begin{figure}
\begin{center}
\includegraphics[width=8.5cm]{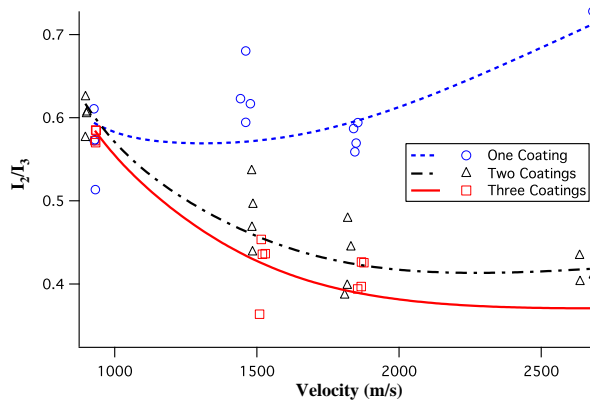}
\caption{(Color online) The ratio of the intensities of the second and third diffraction orders $I_2/I_3$ is shown as a function of velocity for grating G1 at normal incidence. Measurements were made after repeated coating with Au/Pd. The curves indicate three least squares fits to the data.}
\label{fig:coating_i2i3}
\end{center}
\end{figure}

Our experimental setup is described elsewhere \cite{Perreault:2005kx}. In brief, we used a supersonic beam of Na atoms incident on a SiN$_x$ material grating with a period of $d=100$ nm (Fig. \ref{fig:gr_dr}). We measured the flux of diffracted atoms as a function of position by translating a hot-wire detector in the transverse direction (Fig. \ref{fig:raw_data}). We then did a least squares fit of these data to determine the diffraction order intensities ($I_n$). We measured $I_n$ at different grating rotation angles $\beta$, as a function of velocity, ranging from 1000 to 3000 m/s. In the analysis we used only the ratio $I_2/I_3$ vs velocity (Fig. \ref{fig:coating_i2i3}) to determine both $C_3$ and $w$. We choose to focus on these two orders because they are more sensitive to $C_3$ than the 0th and 1st orders and more easily detectable than higher orders. Studying the ratio reduces systematic errors associated with detector non-linearity and the beam-profile used to fit the diffraction data. When measuring diffraction intensities as a function of grating rotation, we recorded a complete diffraction pattern for each rotation angle, thus taking into account the change in separation of the orders due to foreshortening of the grating period; this was ignored in previous work by Cronin \textit{et al.} \cite{Cronin:2004yq}.
 
Given the typical grating geometry, we know all atoms must pass within 30 nm of a surface. We are therefore sensitive exclusively to the short range non-retarded VdW potential. We used the model described in reference \cite{Perreault:2005kx} to fit the diffraction intensities. The potential was approximated by that of two surfaces of infinite extent, co-located with the two inside surfaces of a grating window. The potential of one such a surface is well known to have the form
 
\begin{equation}
\label{eqn:VdWpotential}
V(r) =-  \frac{C_3}{r^3}
\end{equation}

\noindent where $r$ is the distance from the atom to the surface. This potential was then assumed to be `on' only for a distance $t$ while the atom is between the grating bars. In the WKB approximation, the phase of the atom wave-function just beyond the grating is given by

\begin{equation}
\label{eqn:phi_vdw}
\phi_{\textrm{VdW}}(x)=-\frac{1}{\hbar v}\int_{z_0}^{z_0 + t} V_{\textrm{VdW}}(z, x; w, \beta, \alpha, t) \textrm{ d}z  
\end{equation}

\noindent where $z$ is the direction of propagation and $x$ is the the transverse coordinate (see Fig. \ref{fig:gr_dr}). The intensity of the far-field diffraction orders can now be calculated using techniques from Fourier optics.

\section{Magic open-fraction}

The question of whether there is an optimal nano-grating geometry to use in diffraction experiments has been a topic of informal conversation for some time \cite{Kornilov2006}. We will show that there is indeed a ``magic'' open-fraction that improves the precision of $C_3$ measurements by a factor of 5. 
The VdW-induced phase has curvature, just as would be the case for the phase acquired by light due to a lens. Thus we see that the VdW potential causes atoms to be deflected into higher orders.  Similarly, a grating with a smaller window width would lead to more intensity in the higher orders relative to the zeroth order. It is tempting to try to approximate the effect of the VdW potential by a simple diffraction model that uses $C_{3\textrm{eff}}=0$ and a modified window width 
\begin{eqnarray}
\label{eqn:approx}
C_{3\textrm{eff}}&=&0\nonumber\\
w_{\textrm{eff}} &=& w - \Delta w(C_3).
\end{eqnarray}

\noindent  This approximation has become the matter of textbooks \cite{Foot2005}, and it illustrates how a measurement of the VdW potential cannot be made unless the window width is well known. We derive an expression for $w_{\textrm{eff}}$ in Appendix A.

The fact that we can make such an approximation explains the correlation between the fit-parameters $C_3$ and $w$. It is therefore of interest to examine whether or not this approximation holds for all $C_3$ and $w$.  We recall that diffraction from a binary amplitude mask produces zero intensity in the $m$-th order when the open-fraction $w/d=1/m$. However, Cronin \textit{et al.} \cite{Cronin:2004yq} have demonstrated that, in the presence of the VdW potential, diffraction from material gratings will never produce a diffraction order with zero intensity. We therefore expect that there exists a particular magic open-fraction where the approximation in eqn. (\ref{eqn:approx}) breaks down. We expect this to occur when the effective open-fraction approaches an integer fraction, i.e.
\begin{equation}
\label{eqn:magic}
w \rightarrow w_{\textrm{magic}} \qquad \textrm{as} \qquad w_{\textrm{eff}}/d \rightarrow 1/m.
\end{equation}
\noindent In Appendix A we derive an expression for $w_{\textrm{magic}}$ and we show that the correlation between $C_3$ and $w$ indeed vanishes as $w \rightarrow w_{\textrm{magic}}$.

\begin{figure}
\begin{center}
\includegraphics[width=8.5cm]{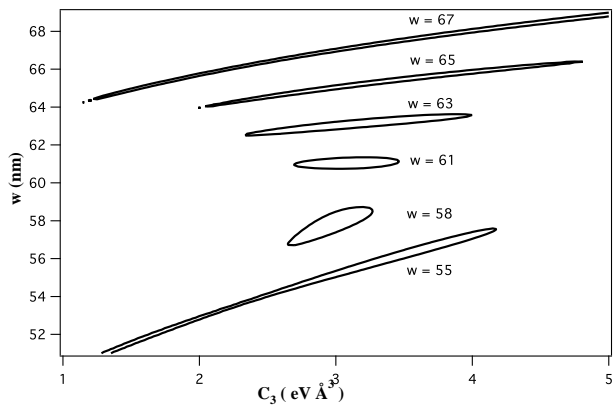}
\caption{The figure shows contours of $\chi^2=\chi^2_{min}+1$ in the $C_3$-$w$ plane corresponding to the 1-$\sigma$ uncertainty of the parameters. Contours are shown for simulated data sets with $C_3=3 \textrm{ eV} \mathring{\textrm{A}}^3$, $d=$ 100 nm, and different values of $w$. The contours describe extended valleys at most open-fractions but narrow to a well defined minimum for one particular magic open-fraction. In this example $w_{\textrm{magic}} =$ 61 nm, in general this value is dependent on $C_3,\alpha$ and $t$, as described in Appendix A.}
\label{fig:chisq_simdata}
\end{center}
\end{figure}

For simplicity, we first use simulated data to investigate what happens to the fit-parameters $C_3$ and $w$ when $w\rightarrow w_{\textrm{magic}}$. We simulated $I_2/I_3$ vs velocity (similar to Fig. \ref{fig:coating_i2i3}) for several grating geometries and fitted the simulated data with our model. Figure \ref{fig:chisq_simdata} shows a contour-plot of the $\chi^2$-surface in $C_3$ - $w$ space for several different values of $w$. For most values of $w$ these contours describe extended valleys, making it impossible to determine either parameter uniquely. Near the magic open-fraction however, $\chi^2$ has a well defined minimum.

To produce a grating with the magic open-fraction we started with a grating that had a larger open-fraction, coated it with a thin layer of metal and measured atom diffraction. We repeated this procedure until the magic open fraction was reached. The samples were coated using a Hummer IV sputter coater with a Au/Pd target. The coating was applied using a plasma current of 2 mA for 30 seconds at a pressure of about 100 mTorr; this nominally corresponds to a 1 nm layer of deposition. The mean free path length under these sputtering conditions is about 1 mm which means that the deposition occurs at a wide range of angles (omnidirectional) and that a simple geometric model of the atomic trajectories suffices to determine how much material is deposited on the inside walls of the grating bars. A coating was applied to both the front and back of the grating.

\begin{figure}
\begin{center}
\includegraphics[width=8.5cm]{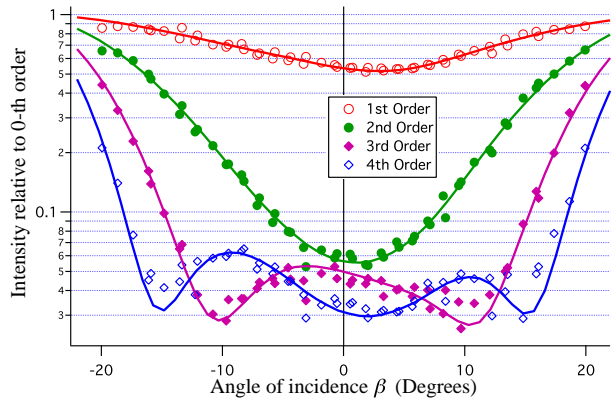}
\caption{(Color online) The intensity of orders 1, 2, 3, and 4 relative to the 0-th order as a function of the grating rotation $\beta$ (see Fig. \ref{fig:gr_dr}). Each set of points at one angle represents one diffraction scan like the one shown in Fig. \ref{fig:raw_data}. A least squares fit to these data allows us to determine $\alpha$ and $\beta_c$ independently from $w$ and $C_3$. Although $w$ and $C_3$ are left as free parameters in this fit, they are not well constrained. The data represented here are for a grating determined to have $\alpha= 5\pm1$ degrees and $\beta_c=26.5\pm0.1$ degrees. For a further discussion of rocking curves see references \cite{Cronin:2004yq,Barwick:2006bf}.}
\label{fig:in_all_2000}
\end{center}
\end{figure}

A precise determination of $C_3$ and $w$ requires additional knowledge of a grating's geometric parameters. We obtain these parameters from rocking curves by doing a least squares fit to the data shown in Fig. \ref{fig:in_all_2000}. A careful analysis of the covariance matrix for this fit shows that there are two parameters, that can be determined independently of $C_3$ and $w$. These parameters are the wedge angle ($\alpha$), and the rotation angle beyond which no atoms can pass through the grating ($\beta_c$).

\begin{figure}
\begin{center}
\includegraphics[width=8.5cm]{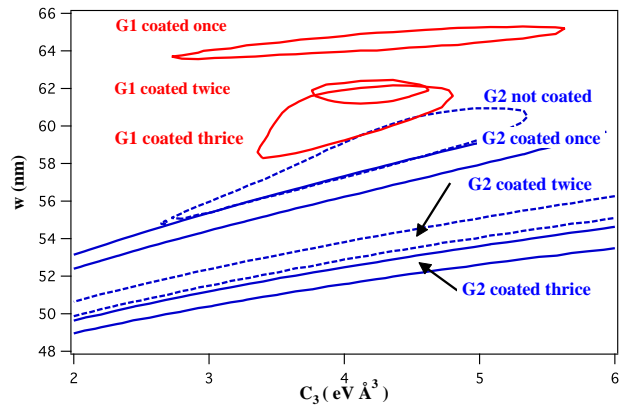}
\caption{(Color online) Contours of $\chi^2=\chi^2_{min}+1$ in $C_3$-$w$ space for two gratings (labeled G1 and G2) after repeated coatings with Au/Pd. The small contour for ``G1 coated twice'' shows the impact of the magic open-fraction.}
\label{fig:coating_contours}
\end{center}
\end{figure}

Armed with this knowledge of $\alpha$ and $\beta_c$,  we measured the ratio $I_2/I_3$ as a function of velocity to determine $C_3$ and $w$. These data are shown in Fig. \ref{fig:coating_i2i3}. We did this several times: after one, two and three coatings for two different gratings. Figure \ref{fig:coating_contours} shows $\chi^2$ contour plots corresponding to fits of these data. The shape of the $\chi^2$ contours match very well to the simulation shown in Fig. \ref{fig:chisq_simdata}. We can see that for a window width a few nm larger or smaller than about 60 nm, the $\chi^2$ contours describe long valleys. The best fit $C_3= 4.2\pm0.4 $ eV$\mathring{\textrm{A}}^3$ is significantly higher than expected for a SiN$_x$ surface but is consistent with a metal coated grating. We will interpret the measured values for $C_3$ sections \ref{sec:pwi} and \ref{sec:metal}.
 
\section{Magic Angle}

In the study of atom-surface interactions it is usually undesirable to cover the surface with any coatings. We can use a technique similar to the one above on a grating of any initial open-fraction, by rotating the grating around a grating bar. When the grating is rotated by an angle greater than the wedge angle $\alpha$, the projected open-fraction is reduced (see Fig. \ref{fig:gr_dr}). When we rotate the grating, the acquired phase profile changes as the inside surfaces of the grating bar are either rotated towards or away from the path of the atoms. Taking into account this change, we can rotate the grating by a particular ``magic'' angle $\beta_m$ such that eqn. (\ref{eqn:magic}) is satisfied.

A grating was placed on a motorized rotation stage with an optical encoder with $1/25$ degree precision. We acquired diffraction scans at about 50 angles ranging from -25 to 25 degrees and fitted the 0-th order intensity as a function of angle for an a posteriori calibration of normal incidence. 

We do not know a priori at what angle to position the grating, since we would need to know both the window width and $C_3$ to calculate the magic angle. We can however detect the magic angle in a diffraction experiment. From eqn. (\ref{eqn:magic}) we expect that the intensity of one of the orders will be minimal at the magic angle. Figure \ref{fig:in_all_2000} shows the intensities of orders 1 through 4 relative to the zeroth order obtained from 50 diffraction scans at different angles. The third order is minimal at an angle of 10.5 degrees. We therefore expect the magic angle to be around 10.5 degrees. 

\begin{figure}
\begin{center}
\includegraphics[width=8.5cm]{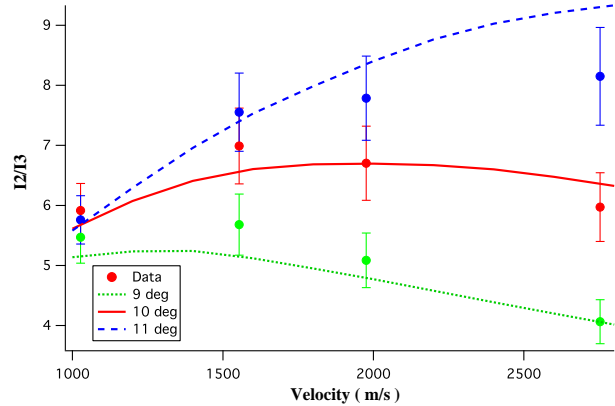}
\caption{(Color online) The ratio $I_2/I_3$ as a function of velocity at three different angles of incidence, near the magic angle. The solid lines indicate a global fit to the sixteen data points that are shown. The data shown here is for the grating labeled G3.}
\label{fig:i2i3}
\end{center}
\end{figure}

\begin{figure}
\begin{center}
\includegraphics[width=8.5cm]{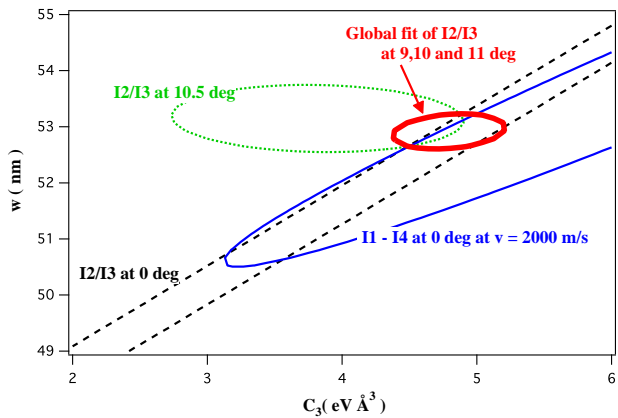}
\caption{(Color online) Contours of $\chi^2=\chi^2_{min}+1$ in $C_3$-$w$ space for the Au/Pd coated grating labeled G3. The blue (thin solid) ellipse corresponds to a fit 
of a single diffraction scan like Fig. \ref{fig:raw_data} at 2000 m/s at normal incidence. The black (dashed) and green (dotted) ellipses correspond to data similar to Fig. \ref{fig:i2i3} at a single angle $\beta=0$ and $\beta=10.5$ degrees respectively. The red (bold solid) ellipse corresponds to the global fit of the three angles shown in Fig. \ref{fig:i2i3} resulting in the best constraint on $C_3$.}
\label{fig:chisq_angledata}
\end{center}
\end{figure}

We did this experiment for two different gratings, one that was pure SiN$_x$ and one that had been coated with Au/Pd. The results for the Au/Pd coated grating are shown in Fig. \ref{fig:i2i3} and \ref{fig:chisq_angledata}. The magic angle was found to occur at 10.5 degrees. To achieve the best precision in $C_3$ and $w$ we did a global fit to $I_2/I_3$ vs velocity at 9, 10 and 11 degrees, shown in Fig. \ref{fig:i2i3}. The combination of $C_3$ and $w$ obtained this way is consistent with diffraction data at normal incidence, as can be seen from Fig. \ref{fig:chisq_angledata}. We find $C_3=4.8 \pm 0.5$ eV$\mathring{\textrm{A}}^3 $, which is once again significantly larger than the value expected for a SiN$_x$ surface.

The same experiment on a grating with no Au/Pd coating on the other hand yields $C_3 = 3.26\pm 0.16$ eV$\mathring{\textrm{A}}^3 $, which is consistent with theory for a SiN$_x$ surface. The $\chi^2$ contour for the clean grating is shown in Fig. \ref{fig:Na_exposure_chisq}. The values found for $C_3$ for different gratings and coatings are summarized in Table \ref{theory_table}. The fit parameters for both gratings are summarized in Table \ref{results}. In order to find the physical $C_3$ for this combination of atom and surface, we need to take into account the shape of the potential more carefully. This is discussed in section \ref{sec:pwi}.

The grating labeled G3 was used by Lepoutre \textit{et al}. \cite{Lepoutre2009} to measure VdW induced phase shifts in an atom interferometer. Our determination of the grating geometry enabled Lepoutre \textit{et al}. to verify the $1/r^3$ form of the VdW potential. 

\section{Theoretical Predictions \label{sec:theory}}

An ab-initio calculation of the the atom-surface interaction can be a very challenging undertaking; even a calculation for idealized atoms and surfaces can only be done analytically in a hand-full of special cases. For an arbitrary geometry there is as of yet no proven method to exactly calculate the interaction strength \cite{Maclay:2001ad}. For a real system, we must consider the frequency response of the atom as well as that of the surface, which is affected by the composition and the geometry of the surface. For an infinite plane surface, the VdW coefficient in the non-retarded regime is given in reference \cite{ZAREMBA:1976ty} by

\begin{equation}
\label{theory_c3}
C_3 = \frac{\hbar}{4 \pi}\int_{0}^{\infty} \alpha_\textrm{pol}(i\omega) \frac{\epsilon(i \omega)-1}{\epsilon(i \omega)+1} \textrm{ d}\omega
\end{equation}

\noindent where $\alpha_\textrm{pol}(i\omega )$ is the atomic polarizability and $\epsilon(i \omega)$ is the dielectric response function of the surface in atomic units. In general $\alpha_\textrm{pol}(i \omega )$ is given by 

\begin{equation}
\label{osum}
\alpha_\textrm{pol}(i\omega)= \sum_{n} \frac{f_n}{\omega_n^2 + \omega^2}
\end{equation}
\noindent where $f_n$ are the oscillator strengths for transitions from the ground state to all the $n$ other states with $\sum f_i=1$. For the present case it is sufficient to include only the two D-lines of sodium since their combined oscillator strength $f_{D1}+f_{D2}  = 0.961$. Indeed, for $\omega=0$ eqn. (\ref{osum}) yields $\alpha_\textrm{pol}(0)= 160.7$  a.u. = 23.81 $ \mathring{\textrm{A}}^3$ which accounts for $99\%$ of the static dipole polarizability of sodium. 

The optical response of the silicon nitride was obtained experimentally in reference \cite{PHILIPP:1973bh}. The material studied in this reference was produced in a way similar to the material of our gratings. The stoichiometry of the SiN$_x$ is not exact, but $x$ is nearly equal to 4/3.

Theoretical calculations by Derevianko \textit{et al.} \cite{Derevianko:1999db} have shown that the effect of core electrons to $C_3$ can be significant. They report a 15\% increase in $C_3$ for sodium and the surface of an ideal conductor as compared to a model not including core electrons.  For a SiN$_x$ or a gold surface, theory predicts a more modest increase in $C_3$ on the order of 6\%. This is understandable since core excitations mainly contribute to $\alpha_\textrm{pol}(i\omega)$ at high frequencies where the response of a real surface is small. Our measurements using a SiN$_x$ surface are in better agreement with the model that includes core-electrons, though the deviation from a simple Lorenz oscillator model is only 1 standard deviation.
The results are summarized in Table \ref{theory_table}.

\section{Pairwise interaction \label{sec:pwi}}

The parameter $C_3$ is only defined for an atom near an infinitely extended plane surface for which the potential has a well known $C_3/r^3$ dependence. In the case of a more complicated geometry such as our gratings, the potential is in principle a function of all three spatial coordinates $\left(V_\textrm{VdW} = V_\textrm{VdW}(x,y,z)\right)$. There are multiple ways of defining a $C_3$ in this case. One way is the approximation discussed before eqn. (\ref{eqn:VdWpotential}), where we use the potential of two infinite planes co-located with the grating's inner walls. This approximation however ignores the effect of edges. A more precise way to define $C_3$ is to use the pairwise interaction (PWI) model.

The PWI approximation assumes that the interaction of an atom with a solid body is proportional to the sum of the interactions with each of the atoms composing the body. Mostepanenko \textit{et al.} suggested that the PWI model, which neglects any multi-body interactions and screening effects, gives the right spatial dependence of the potential, but requires a normalization constant in order to yield the correct potential \cite{MOSTEPANENKO:1988ty}. The PWI potential is given by

\begin{equation}
\label{eqn:pwi}
V(\mathbf{r}) =- K \int_V \textrm{d}^3 r' \frac{n C_6}{|\mathbf{r-r'}|^6}
\end{equation}

\noindent where the normalization constant $K$ is dependent on the geometry of the solid, and its material properties, $n$ is the number-density of atoms and $C_6$ is the atom-atom interaction constant. 

The normalization constant can be obtained by comparing the PWI model with an exact calculation for an infinite plane. We can then express $K n C_6$ in terms of $C_3$ for an infinite plane:

\begin{eqnarray}
V &=&- \int_{x>0}\textrm{d}x \int\textrm{d}y \int\textrm{d}z  \frac{K_{\textrm{plane}} n C_6}{[(x+d)^2 +y^2 + z^2]^{6/2}}\\
&=& -\frac{K_{\textrm{plane}} \pi n C_6}{6 d^3}\\
&=& -\frac{C_3}{ d^3}
\end{eqnarray}

\noindent where $d$ is the distance to the surface. We can now identify $C_3 = \pi K_{\textrm{plane}} n C_6/6$. A numerical evaluation of the integral in eqn. (\ref{eqn:phi_vdw}) using the potential in eqn. (\ref{eqn:pwi}) with $ K n C_6=6 C_3 / \pi $, shows that the the PWI model yields a smaller $\phi_{VdW}$ by about 5\% as compared to the approximation discussed before eqn. (\ref{eqn:VdWpotential}). We accordingly apply a  5\% correction to the $C_3$ found from our fits.  

By construction the PWI model gives the correct potential near the middle of a grating bar and close to the surface, where the effect of edges is negligible. Near an edge, the PWI model may yield a potential that differs from the actual potential by no more than 13\%. We obtained this upper bound to the error by comparing the PWI model to the exact solution in the limiting case of an atom far away from a perfectly conducting sphere \cite{CASIMIR:1948cl,Buhmann:2004fe}.  In this case the PWI potential is expected to differ most from the real potential by reasoning analogous to that in reference \cite{MOSTEPANENKO:1988ty}. We further assume that the real potential of an atom near a surface of finite extent must be smaller than the potential near an infinite surface thus creating an additional upper bound to the error. Based on an analysis of the error along the propagation path of the atom, we conclude that the PWI correction introduces an error that is no larger than 3\% to the acquired phase, and thus to our measurement of $C_3$.

\begin{table*}[htdp]
\caption{Experimental results and theoretical predictions}
\begin{center}
\begin{ruledtabular}
\begin{tabular}{ l | c  l }

Experiment & $C_3$ (eV$\mathring{\textrm{A}}^3 $) &  \\
\hline
This work:&&\\
Na and SiN$_x$ fit parameter, grating A &3.26& $\pm$ 0.16 \footnotemark[1] \\
Na and Au/Pd fit parameter, grating G1&4.30& $\pm$ 0.5\\
Na and Au/Pd fit parameter, grating G3&4.80& $\pm$ 0.5\\
&&\\
Na and SiN$_x$ with PWI correction, grating A&3.42& $\pm$ 0.19 \footnotemark[1]\\
Na and Au/Pd with PWI correction, grating G1&4.51& $\pm$ 0.5\\
Na and Au/Pd with PWI correction, grating G3&5.04& $\pm$ 0.5\\
\hline
Previous work:&&\\
Na and SiN$_x$ \cite{Perreault:2005kx}&2.70&$\pm$ 0.8\\

\hline
\hline
Theory&&\\
\hline

Na and perfect conductor&7.6&\cite{Derevianko:1999db}\\
Na and SiN$_x$ using single oscillator $\alpha_\textrm{pol}(i\omega)$ and tabulated $\epsilon(i\omega)$&3.3&\cite{PHILIPP:1973bh}, eqn. (\ref{theory_c3}) \\
Na and SiN$_x$ using tabulated $\alpha_\textrm{pol}(i\omega)$ and $\epsilon(i\omega)$&3.48&\cite{PHILIPP:1973bh,Derevianko2009}, eqn. (\ref{theory_c3})\\
Na and Bulk Au&5.11& \cite{Caride:2005tg}\\
Na and 1 nm Au at $r=$ 10 nm&4.3&\cite{ZHOU:1995sf}\\
Na and 2 nm Au at $r=$ 10 nm&4.5&\cite{ZHOU:1995sf}\\
Na and 3 nm Au at $r=$ 10 nm&4.6&\cite{ZHOU:1995sf}\\

\end{tabular}
\end{ruledtabular}
\end{center}
\footnotetext[1]{The sources of the reported uncertainty are discussed in Appendix B.}
\label{theory_table}
\end{table*}

\section{The effect of a thin layer of metal \label{sec:metal}}

The two measurements of $C_3$ for Na and an Au/Pd coated grating both showed a significantly larger $C_3$ than is expected for a clean SiN$_x$ surface. The magic open-fraction experiment and the magic angle experiment were done using two different samples and give $C_3 =4.3 \textrm{ eV} \mathring{\textrm{A}}^3$ and $C_3 =4.8 \textrm{ eV} \mathring{\textrm{A}}^3$ respectively. These results are in agreement with theoretical predictions for $C_3$ near a gold coated surface.

A thin surface layer (roughly less than an optical skin-depth) does not produce the same VdW potential as a bulk material, even in the idealized case of a uniform layer that retains the optical response properties of the bulk material. Studies of the Casimir potential between large bodies have detected a dependence on surface properties at distances greater than 100 nm \cite{Lisanti:2005rt}. For thin layers, less than a skin-depth of the surface material, the effect of the surface layer is significantly reduced at these distances. Based on theoretical work however, we expect the effect on the short range VdW potential to be significant even for very thin layers \cite{ZHOU:1995sf}.   

Our metal coated gratings consist of a thin coating of a Au/Pd mixture on top of a silicon nitride nano-bar.  
In modeling such a system we must take into account the interface between the vacuum and the thin outer surface as well as the interface between the surface layer and the underlying bulk material. To do this, we have evaluated expression 4.14 in reference  \cite{ZHOU:1995sf}. 
We used a Drude model for $\epsilon_{\textrm{Au}}(i \omega)$ and an insulator model for $\epsilon_{\textrm{SiN$_x$}}(i \omega)$. 

Two problems present themselves in a precise interpretation of the results. First, the atom-surface potential is dependent on the thickness of the coating, which is not very well known and may be non-uniform. Second, in the presence of a thin surface layer the VdW potential no longer follows an exact $1/r^3$ potential. For a 2 nm layer of Au, $C_3$ varies from 4.8 eV$\mathring{\textrm{A}}^3$ to 4.2 eV$\mathring{\textrm{A}}^3$ at distances of 2 nm to 25 nm from the surface respectively. Where we used $C_3 = V(r)\cdot r^3$  
We evaluated the potential at a distance of 10 nm, where our experiments are most sensitive \cite{Perreault:2005kx}. A 2 nm layer of Au on the surface gives $C_3=4.5$ eV$\mathring{\textrm{A}}^3$. This value depends on the thickness of the Au layer, a 3 nm layer for example gives $C_3=4.6$ eV$\mathring{\textrm{A}}^3$. 

In the analysis of our data, we still assumed a $1/r^3$ potential since we are most sensitive to a small range of distances and the physical potential only deviates weakly from this form over a short range. Any errors introduced by this assumption must be smaller than the range of $C_3$ given above and are of the order of the uncertainty reported. 

Despite these difficulties, the reported $C_3$ for a metal coated grating is significantly different from the SiN$x$ measurement. The change of more than 30\% is consistent with theory (see Table \ref{theory_table}). Our measurements on both metal coated samples are clearly inconsistent with a clean SiN$_x$ surface.

The growth of thin films of metal is a complicated process subject to many experimental parameters \cite{Goldstein2003}, however there are several pieces of evidence that show that there is a significant amount of metal on the inside surfaces of the grating. First, we know that the mean free path length in the coating process is about 1 mm. This means the deposition is omnidirectional and the inside of the bars will be coated with nominally 1 nm of metal. Second, SEM images of the metal coated grating show significantly reduced effects of charging as compared to before coating, this indicates the surface is conducting charge. Third, x-ray spectra we obtained in the SEM show the presence of Au. Finally, the measured $C_3$ is consistent with metal coating on the inside of the bars. It is however inconsistent with both theoretical predictions and our measurement of $C_3$ for a clean SiN$_x$ grating. 

\section{The effect of the Na beam on the grating}

During an experiment, a grating is exposed to the atom beam for several hours. A set of 50 diffraction scans used to 
make Fig. \ref{fig:in_all_2000} takes about 30 minutes to acquire. Such a data-set is taken at four different 
velocities to fully characterize a grating. The total exposure time to the atom beam is about 2 hours per experiment. To determine if prolonged beam exposure causes contamination and a affects $C_3$, we exposed the grating to a high flux atom 
beam. Under normal experimental conditions the atom beam is collimated by two $10\textrm{ } \mu$m wide slits which attenuate the 
flux at the location of the grating by a factor of about 100. To test the effect of beam exposure we removed both collimating slits and 
exposed the grating to the unattenuated atom beam source for a duration of 2 hours. 

\begin{figure}
\begin{center}
\includegraphics[width=8.5cm]{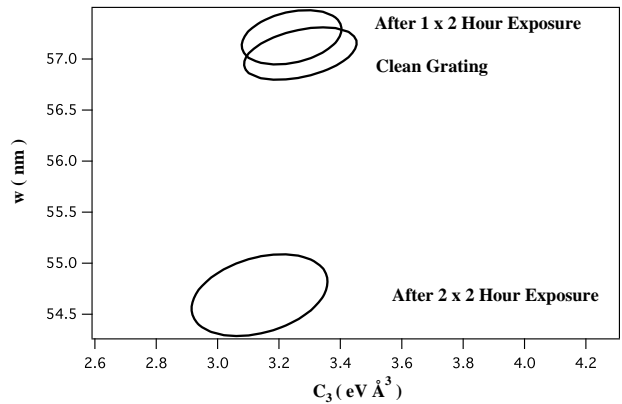}
\caption{The effect of exposure to the Na atom beam is demonstrated. One 2 hour exposure in this figure corresponds to a dose 
100 times larger than in a typical experimental run. The clean SiN$_x$ grating shows a $C_3$ of $3.26 \pm 0.16$ eV$\mathring{\textrm{A}}^3 $ }
\label{fig:Na_exposure_chisq}
\end{center}
\end{figure}

The results are shown in Fig. \ref{fig:Na_exposure_chisq}. After the first 2 hours, the open-fraction is not significantly affected, after another 2 hour exposure, the window width is changed by 2.5 nm. Table \ref{results} shows the best fit geometric parameters before and after coating, A$_0$ represents a clean grating, A$_1$ and A$_2$ correspond to the same grating after one and two exposures respectively. 

The best fit value for $C_3$ is not significantly changed after exposure to the atom beam. We propose this is due to non-uniform coating of the grating. The directionality of the atom beam suggest that the majority of the atoms would hit the front face of the grating bars, only about 10\% would hit the side of the bars directly. AFM images confirm that the sodium forms large clumps on the front face of the grating that overshadow the grating windows.  The change in the wedge angle after the coating is consistent with this explanation.

We conclude that under normal experimental conditions, the effect of beam exposure should be 100 times smaller corresponding to 0.01 nm 
per 2-hour experiment. These deposition rates are consistent with the atom count rate we measure at the detector. A small amount of contamination will still be present.

\begin{table}[htdp]
\caption{Best fit parameters found from least squares fits for  three different gratings. The origin of the quoted uncertainties is explained in Appendix B.  A$_0$ represents a clean grating, A$_1$ and A$_2$ correspond to the same grating after one and two exposures to the Na beam respectively. G1 and G3 are Au/Pd coated gratings. G3 is the Au/Pd coated interaction grating used in reference \cite{Lepoutre2009}.}
\begin{center}
\begin{ruledtabular}
\begin{tabular}{ c| l l | l l | l l | l l }
Grating& $C_3$& $(\textrm{eV} \mathring{\textrm{A}}^3)$&$w$ 	&	(nm)		&$\alpha $		&	(deg)	& $t$  &(nm)\\ 
name&&&&&&&&\\
\hline
&&&&&&&&\\
A$_0$\footnotemark[1]\footnotemark[3]		&3.26	&$\pm$ 0.16	&57.0	&$\pm$ 1.0	&3.5		&$\pm$ 0.5	& 140 	&$\pm$ 10\\
A$_1$\footnotemark[1]\footnotemark[4]		&3.24   	&$\pm$ 0.3 	&57.5	&$\pm$ 1.0	&3.3		&$\pm$ 0.5	& 140 	&$\pm$ 10\\
A$_2$\footnotemark[1]\footnotemark[4]		&3.1     	&$\pm$ 0.3 	&54.5	&$\pm$ 1.0	&3.1		&$\pm$ 0.5	& 140 	&$\pm$ 10\\
G1\footnotemark[2]\footnotemark[5]		&4.3		&$\pm$ 0.5	&61.6	&$\pm$ 1.0	&4.5		&$\pm$ 0.5	& 110 	&$\pm$ 10\\
G3\footnotemark[1]\footnotemark[5]		&4.8   	&$\pm$ 0.5 	&53.0	&$\pm$ 1.2	&5.0		&$\pm$ 1.0	& 110 	&$\pm$ 10\\

\end{tabular}
\end{ruledtabular}
\footnotetext[1]{$C_3$ and $w$ determined by magic angle method.}
\footnotetext[2]{$C_3$ and $w$ determined by magic open-fraction method.}
\footnotetext[3]{Clean SiN$_x$ grating.}
\footnotetext[4]{Grating exposed to Na atom beam.}
\footnotetext[5]{Au/Pd coated grating}
\end{center}
\label{results}
\end{table}

\section{Conclusion}

We measured the VdW potential for Na and a SiN$_x$ surface with 5 times better precision than previous work. We made use of the fact that using a grating with a particular magic open-fraction or a magic rotation angle, $C_3$ can be determined independently from the geometric grating parameters. This method also yields a precise determination of the geometric parameters of the grating. These measurements are not subject to the systematic problems that plague conventional imaging techniques and are therefore useful in other experiments using nano-gratings. 
Our measurements are precise enough to detect an increase in the atom-surface potential due to a thin layer of metal. We also detected the effect of extended exposure to the atom beam on the grating parameters, however the effect of contamination by the atom beam was verified to be negligible under normal experimental conditions. Our measurements are now approaching a precision where they are sensitive to edge-effects and the contribution of core electrons.
\\
\\
This material is based upon work supported by the National Science Foundation under Grant No. PHY-0653623 and the Arizona TRIF imaging program. The authors thank the members of the Vigu\'e team in Toulouse and Oleg Kornilov and J. Peter Toennies for many fruitful discussions.

\appendix
\section{Deriving the magic open-fraction}

\begin{figure}
\begin{center}
\includegraphics[width=8cm]{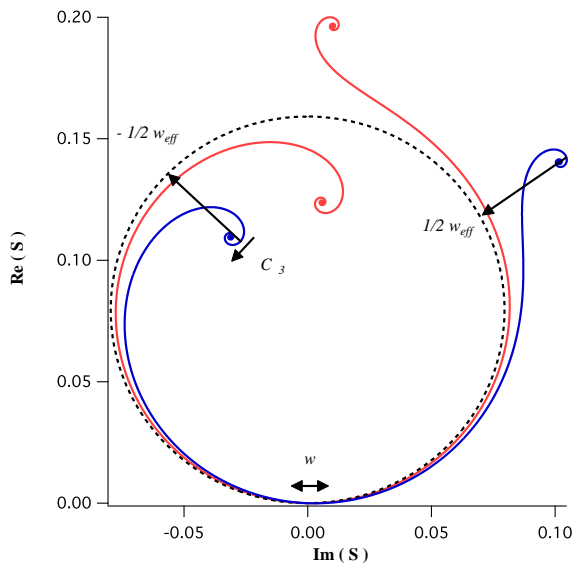}
\caption{(Color online) A Cornu spiral is a graphical representation of the integral in eqn. (\ref{eq:Phi_n}), it is a curve in the complex plane parameterized by $\{\Re[S(q)],\Im[S(q)]\}$ where $S(q) = \frac{1}{d}\int_{-w/2}^{q}\exp\left[i \left(-\frac{2 \pi n x}{ d}+\phi_{VdW}(x)\right)\right] \textrm{ d}x$. The length of the curve is equal to the domain of integration ($w$), while the distance between the endpoints gives the absolute value of the amplitude of the diffraction order. The figure shows a Cornu spiral in blue (dark grey) for the second order ($n$=2) for a grating with $w$ = 50 nm and $C_3=3 \textrm{ eV} \mathring{\textrm{A}}^3$. The figure also shows a Cornu spiral for a grating with the magic open-fraction in red (light grey).}
\label{fig:cornu}
\end{center}
\end{figure}

To better understand the correlation between the window width and the VdW coefficient, we consider diffraction from a grating with rectangular bars. In the far field, the intensity of the n-th order is $I_n = |A_n|^2$. Where the complex amplitude $A_n$ is shown in reference \cite{Perreault:2005kx} to be
\begin{equation}
\label{eq:Phi_n}
A_n=\frac{1}{d}\int_{-w/2}^{w/2}\exp\left[i \left(-\frac{2 \pi n x}{ d}+\phi_{VdW}(x)\right)\right] \textrm{ d}x
\end{equation}
\noindent with 
\begin{equation}
\phi_{VdW}(x)=\frac{C_3 t}{\hbar v} \left( \frac{1}{(w/2-x)^3} + \frac{1}{(w/2+x)^3}\right)
\end{equation}

\noindent where $w,d$ and $t$ are the grating window width, period and thickness, $v$ is the velocity of the atoms and $x$ is the position along the grating window (Fig. \ref{fig:gr_dr}). This integral is represented graphically in Fig. \ref{fig:cornu} by a Cornu spiral \cite{Halliday2002,Cronin:2004yq}. The length of the curve is equal to the window width, while the distance between the endpoints gives $|A_n|$. For $C_3=0$ the curve would lie on the circle shown in the figure; when $C_3$ is non-zero the extremities of this curve spiral away from the circle, one inside the circle and one outside.  This immediately demonstrates that in the presence of the VdW potential, there are no orders with zero intensity. 

To study the effect of small changes $\Delta C_3$ and $\Delta w$ on the diffraction intensity, we need to look at the corresponding translation of the endpoints. As indicated in the figure, a change in $w$ adds a length in the middle of the curve where the curvature of the Cornu spiral is nearly equal to that of the circle, thus causing the endpoints of the spiral to move parallel to the circle. A change in $C_3$ increases the curvature of the end of the spiral, which has an effect on the endpoints that is similar but not exactly the same.

To equate a change in $C_3$ to an equivalent change in $w$, we consider only the projection of the endpoint onto the circle that corresponds to $C_3=0$. This neglects radial translations of the endpoints due to $\Delta C_3$. The points on this circle nearest to the end points we label $\pm w_{\textrm{eff}}/2$ for reasons that will soon become clear. To approximate the location of these two points we can use the fact that the part of the spiral where $\phi_{VdW}>\pi$ can be neglected. At the point on the Cornu spiral where $\phi_{VdW}= \pi$, the tangent line of the spiral is antiparallel to the circle. The value of $w_{\textrm{eff}}$ can then be found from eqn. (\ref{eq:Phi_n}) by solving

\begin{equation}
\textrm{exp}\left[i\pi -i\frac{2 \pi n }{ d}x' \right] = - \textrm{exp}\left[-i\frac{2 \pi n }{ d}w_{\textrm{eff}}/2\right]
\end{equation}
\noindent where $x'$ is the coordinate where $\phi_{VdW}= \pi$. The minus sign on the right hand side accounts for the fact that the tangent vectors on the circle ($C_3=0$) and on the Cornu spiral are pointed in opposite directions. The value of $x'$ can easily be found by considering the potential of a single wall, it is however more convenient to define $x_0 = w/2 - x'$, the distance from the atom to the nearest wall. We find $x_0$ by solving

\begin{equation}
\frac{t }{ \hbar v}\frac{C_3}{ x_0^3} = \pi
\end{equation}
 \noindent The relationship between $\Delta w$ and $\Delta C_3$ is then seen to be

\begin{equation}
\Delta w=\frac{\partial w_{\textrm{eff}}}{\partial C_3} \Delta C_3= 2 \frac{\partial x_0}{\partial C_3} \Delta C_3
\end{equation}
\noindent Using common values for the parameters, $t=120$ nm, $v = 1000$ m/s and $C_3 = 3 \textrm{ eV} \mathring{\textrm{A}}^3 $ we get 

\begin{equation}
\frac{\Delta C_3}{\Delta w}=\left(\frac{C_3^{-2} t}{\pi \hbar v}\right)^{-1/3}    \approx 0.8 \frac{\textrm{ eV} \mathring{\textrm{A}}^3}{\textrm{nm}}.
\end{equation}
\noindent This is very close to the relationship found empirically in reference \cite{Perreault:2005kx} between the best fit value for $C_3$ and the fixed value used for $w$. 

A rough approximation for $I_n$ can now be found by setting $C_3=0$ and using $w_{\textrm{eff}}$ in place of the physical value for the window width. This approximation is not very useful in interpreting experimental data but it does allow us to make an estimate of the magic open-fraction. This approximation predicts a zero in the single slit diffraction envelope at order $n$ when $w/d = m/n$ for integer $m$. When $\{n,m\}=\{2,1\}$ the second order is predicted to be missing. Since there are no diffraction orders with zero intensity, we find the magic open-fraction  by solving 

\begin{equation}
\label{eq:weff}
w_{\textrm{eff}}/d = (w_{\textrm{magic}} - 2 x_0)/d = 1/2
\end{equation}
\noindent Using $C_3\approx 3 \textrm{ eV} \mathring{\textrm{A}}^3$, we find $2 x_0\approx 10$ nm  so we predict a magic open-fraction for $w=60$ nm.  A numerical calculation shows that the covariances of the fit parameters $w$ and $C_3$ indeed become small near $w=60$ nm. There are also magic open-fractions near $w=45$ nm and $w=90$nm corresponding to $\{n,m\}= \{3,1\}$ and $\{2,2\}$. In general
\begin{equation}
\frac{w_{\textrm{magic}}}{d} =  \frac{m}{n} + 2\left(\frac{C_3 t}{\pi \hbar v }\right)^{1/3}.
\end{equation}

Figure \ref{fig:cornu} also shows a spiral for a grating with the magic open-fraction, demonstrating how the model using only $w_{\textrm{eff}}$ gives the wrong result.

Near the magic open fraction, the approximation of using an effective open fraction with $C_3=0$ is no longer valid. There is no longer a simple analytic expression for the relationship between the fit parameters $C_3$ and $w$. 
In order to still quantitatively explore $\Delta C_3 / \Delta w$ near the magic open fraction, we have numerically computed this derivative.

\begin{figure}
\begin{center}
\includegraphics[width=8.5cm]{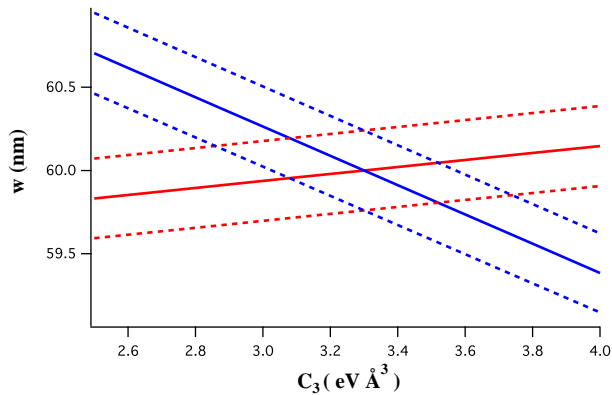}
\caption{(Color online) The two solid lines correspond to the relationship between the best fit $C_3$ and $w$ for a measurement of $I_2$ at velocities $v_1 =$ 1000 m/s and $v_2=$2000 m/s respectively. The dashed lines correspond to the $1\sigma$ confidence interval. The area where the confidence intervals overlap is smallest when the angle $\gamma$ between the two lines is closest to 90 degrees. }
\label{fig:lines}
\end{center}
\end{figure}

We can now show that the correlation between the fit parameters $C_3$ and $w$ indeed vanishes (i.e. $\Delta C_3/\Delta w=0$) at this magic open-fraction. For a single measurement at a single velocity $I_n(v_1)$, we find a relationship between $C_3$ and $w$ that is, at least locally, linear with slope $a_1 = \partial_{C_3}I_n/\partial_{w}I_n$. This relationship is shown in Figure \ref{fig:lines} by the solid line. If we include the uncertainty associated with $I_n$ the linear relationship becomes a confidence region, given by the area between the dashed lines. In order to determine the combination $\{C_3,w\}$ we need another measurement at a different velocity $I_n(v_2)$. If the slope $a_1$ is similar to $a_2$ the confidence region is still an extended valley. Only if the angle between the two lines is large, are both parameters $C_3$ and $w$ well constrained. In that case, as indicated in Fig. \ref{fig:lines}, X marks the spot. From Fig. \ref{fig:lines} we can also see that the condition of minimum correlation is not exactly the same as the condition for the best constraint on $w$ and $C_3$. The former only requires $a_1=-a_2$ while for the latter we want the angle between the two lines to be 90 degrees \cite{Press1992}. The fact that  $\partial_{C_3}I_n/\partial_{w}I_n$ can be both positive or negative means that a suitable combination of measurements can yield  $\Delta C_3/\Delta w\equiv\textrm{Cov}(C_3,w)/\textrm{Var}(w)=0$ for the two fit-parameters. This quantity can be also be obtained by fitting a set of data with $w$ fixed and determining $C_3$ for different values of $w$ \cite{bevington1992}.

\begin{figure}
\begin{center}
\includegraphics[width=8.5cm]{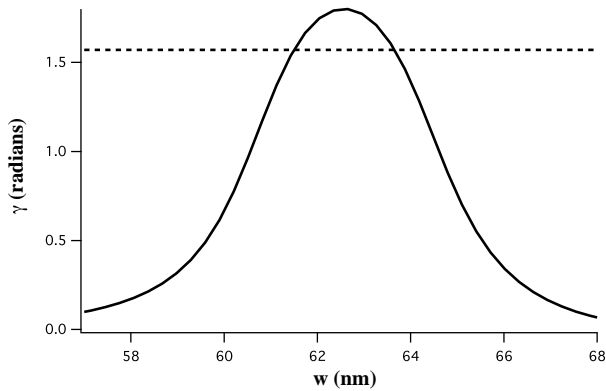}
\caption{The angle $\gamma$ between the two lines in Figure \ref{fig:lines} is plotted versus $w$. The angle is closest to $\pi/2$ around $w=61$ nm which is indeed around where we expect the magic open-fraction to be. As the angle approaches $\pi/2$ the parameters $C_3$ and $w$ become maximally constrained. $\gamma=\pi/2$ is indicated by the dashed line. }
\label{fig:angle}
\end{center}
\end{figure}

The angle between two lines with slopes $a_1$ and $a_2$ is given by  $\gamma = \textrm{atan}(a_1) - \textrm{atan}(a_2)$. Figure  \ref{fig:angle} shows $\gamma$ as a function of $w$ and shows that the angle indeed approaches 90 degrees around the predicted open-fraction $w_\textrm{magic}/d$. The largest $\gamma$ occurs when one of the slopes is negative. This only happens in a narrow range near $w_\textrm{magic}$, when either $\partial_{C_3} I_n $ or $\partial_{w} I_n $ changes sign. We therefore expect the best constraint on $C_3$ to be obtained with an open-fraction (or grating rotation angle) that minimizes $I_n$, as our simple guess predicted.

\section{Discussion of reported uncertainties}

Given the small uncertainty in the reported values for $C_3$, it is appropriate to discuss the various sources of statistical and systematic errors. 
Errors due to surface roughness or variations in the grating parameters have previously been discussed in the literature. First, we consider variations in the grating parameters such as the bar thickness. SEM images that show that the RMS deviation of the grating bar width is about 1.4 nm. We do not observe the exponential dampening predicted in reference \cite{Grisenti:2000ty} so we think surface variations can not be described in terms of a distribution of the geometric parameters. This is consistent with findings in reference \cite{Bruhl:2002ty}. Instead they can be treated as surface roughness.

In quantum reflection experiments in particular, surface roughness influences the reflection efficiency. In our experiment however, surface variations on a scale smaller than the grating bar thickness tend to average out since the atoms sample multiple areas of the surface. The potential near a rough surface, to first order has the same spatial dependence and the same $C_3$ as a smooth surface but relative to an effective surface distance \cite{Maradudin1980}. This should be taken into account in the interpretation of our report of geometric parameters, but is of less significance to our reported values of $C_3$.

We determine the velocity of our atom beam by measuring the distance between diffraction orders. The uncertainty in the velocity is due to the uncertainty in the distance between the grating and the detector which we know to within 1\%. 

As can be seen from Table \ref{errors}, the dominant source of uncertainty is the uncertainty in determining the $I_2/I_3$. This is due to the fact that one of the orders is very small near the magic open-fraction or magic angle.  
The total error given by the individual errors added in quadrature, is 5.7\%.

The PWI correction factor of 5\% was obtained in a simplified geometry (using bars with rectangular cross-section) by computing the correction to the integrated phase (eqn. (\ref{eqn:phi_vdw})) using the PWI potential. We did this for multiple paths through the grating at normal incidence. Since this is a computationally intensive procedure it was not feasible to fit our experimental data with this model and get more exact knowledge of the effect of the PWI correction on the fit-parameters. 

\begin{table*}
\caption{Sources of errors for measurement of $C_3$ for a Na atom and a SiN$_x$ surface.}
\begin{center}
\begin{ruledtabular}
\begin{tabular}{l|l|l|r}

Quantity & Value & Uncertainty& Uncertainty in  $C_3$\\
\hline
Grating-detector distance\footnotemark[1] &2.43 m&1\%&1.0\%\\
Grating wedge angle $\alpha$ &3.5 deg&0.5 deg&0.9\%\\
Critical angle $\beta_c$ &26.5 deg&0.1 deg&1.5\%\\
Geometry variations &2 nm RMS max& 2 nm &1.0\%\\
Surface roughness &2 nm RMS max& 2 nm&1.0\%\\
$I_2/I_3$ from fit of diffraction data& 1 - 7 & 10\%& 4.2\% \\
\hline
Total uncertainty in fit parameter $C_3$  & 3.26$\textrm{ eV} \mathring{\textrm{A}}^3$&0.16$\textrm{ eV} \mathring{\textrm{A}}^3$&4.9\%\\
\hline
PWI correction & 5\% shift & 3\% &3.0\%\\
\hline
\hline
Physical $C_3$&3.42 $\textrm{ eV} \mathring{\textrm{A}}^3$ &0.19 $\textrm{ eV} \mathring{\textrm{A}}^3$&5.7\%\\

\end{tabular}
\end{ruledtabular}
\end{center}
\footnotetext[1]{The grating-detector distance is required to determine the velocity.}
\label{errors}
\end{table*}%

The main source of uncertainty in the determination of $w$ is the uncertainty in the wedge angle $\alpha$. The contours shown in Figures \ref{fig:chisq_angledata} and \ref{fig:Na_exposure_chisq} show a far more stringent bound on $w$ than we ultimately quote in Table \ref{results} because the $\chi^2$-contours correspond to a fit with $\alpha$ fixed. We note that the parameter $w$ refers to the average window width so even though we allow for variations in the physical window width from point to point on the order of 2 nm, the parameter $w$ can be constrained to within 1 nm.


\end{document}